\documentclass[a4,journal]{IEEEtran}
\usepackage[ruled]{algorithm2e}
\usepackage{cite}
\usepackage{graphics}
\usepackage{array}
\usepackage{color}
\usepackage{amsfonts}
\usepackage{import}
\usepackage{amsthm,amsmath}
\allowdisplaybreaks

\usepackage{xpatch}
\makeatletter
\xpatchcmd{\@thm}{\thm@headpunct{.}}{\thm@headpunct{}}{}{}
\makeatother
\usepackage{multirow}
\newcolumntype{L}{>{\centering\arraybackslash}m{0.7cm}}
\graphicspath{{figures/}}
\usepackage{graphicx, pifont} 
\SetAlgorithmName{Model}{model}{list of Models}
\newcommand{\RN}[1]{%
  \textup{\uppercase\expandafter{\romannumeral#1}}%
}

\newcommand{\norm}[1]{\left\|{#1}\right\|}

\begin{document}

\title{Gaussian Process Learning-based Probabilistic Optimal Power Flow}

\author{Parikshit~Pareek,~\IEEEmembership{Student~Member,~IEEE,}~and~
        Hung~D.~Nguyen$^\star$~\IEEEmembership{Member,~IEEE }\vspace{-20pt}
% <-this % stops a space
\thanks{$^\star$Corresponding Author}% <-this % stops a space
\thanks{Authors are with School of Electrical and Electronics Engineering, Nanyang Technological University, Singapore. \textit{pare0001,hunghtd@ntu.edu.sg}}}%

\vspace{-0.3em}

\markboth{IEEE TRANSACTIONS ON POWER SYSTEM Vol. XX}%
{}

% \IEEEaftertitletext{\vspace{-1\baselineskip}}

\maketitle

\begin{abstract}
 In this letter, we present a novel Gaussian Process Learning-based Probabilistic Optimal Power Flow (GP-POPF) for solving POPF under renewable and load uncertainties of arbitrary distribution. The proposed method relies on a non-parametric Bayesian inference-based uncertainty propagation approach, called Gaussian Process (GP). We also suggest a new type of sensitivity called Subspace-wise Sensitivity, using observations on the interpretability of GP-POPF hyperparameters. The simulation results on 14-bus and 30-bus systems show that the proposed method provides reasonably accurate solutions when compared with Monte-Carlo Simulations (MCS) solutions at different levels of uncertain renewable penetration as well as load uncertainties, while requiring much less number of samples and elapsed time.   
\end{abstract}

% Note that keywords are not normally used for peer review papers.

\IEEEpeerreviewmaketitle

\vspace{-0.5em}
\section{Introduction}
\IEEEPARstart{T}{he} upsurge in the renewable penetration and dynamic loads has made \textit{Probabilistic Optimal Power Flow (POPF)} a necessary tool providing necessary uncertainty description in the decision and state variables \cite{sun2019data}. Existing POPF methods fall under analytical, approximate, and Monte-Carlo Simulation (MCS) based categories \cite{sun2019data}. These methods have various limitations, like dependencies on approximate power flow formulations, complicated implementation, or lacking data-based guarantees and substantial sample set requirements. Further, mostly these methods are developed to handle specific input uncertainty distribution such as normal, beta, etc. This becomes a bottleneck in case of limited uncertainty information or when a random variable does not follow any such distribution. The second case is prevalent with Solar Photo-Voltaic (PV) based renewable generation and Electric Vehicle load as their uncertainty forecasting remains a challenge.

In this letter, we introduce a novel POPF solution method that relies on the Bayesian inference based uncertainty propagation technique, Gaussian Process (GP). The GP is used extensively as a supervised learning tool in various machine learning applications \cite{williams2006gaussian}. The proposed \textit{Gaussian Process Learning-based Probabilistic Optimal Power Flow (GP-POPF)} method is build to handle main issues of uncertainty description requirement and a large number of sample requirements of data-based methods. The GP-POPF can be employed to handle arbitrary input uncertainty as, during the training of GP models, specific details of uncertainty distribution are not required. Further, GP provides mean and variance information for a random variable at testing without requiring the distribution information of the random input. As the GP's foundation is the Bayesian inference, the associated interpretability makes GP an ideal candidate for physical network learning like in the proposed GP-POPF. To the best of our knowledge, the GP has not been explored for the task of uncertainty propagation in OPF yet. 

The main contributions of this letter can be summarized as: 
\begin{itemize}
    \item Developing a novel GP-POPF which is a non-parametric in nature, thus be free from a need of pre-defining class of uncertainties. 
    \item Developing the foundation of interpretability in POPF by defining covariance function hyperparameter based on the newly introduced concept of \textit{Subspace-wise Sensitivity}. 
\end{itemize}

We introduce the basic Gaussian Process Regression below.
\vspace{-2.6em}
\subsection{Gaussian Process Regression (GPR)}
The GPR is a non-parametric modeling method providing some level of interpretability of models and allowing to model prior understanding in the data-based models \cite{williams2006gaussian,pareek2019probabilistic}. The GPR interpretability means that upon learning, the covariance matrix contains subtle information about the function it was destined to learn. This information can be then utilized to predict the function  behavior inside as well as outside of the input domain within a limited range. Further, as GPR allows us to provide prior distribution, understanding of the physical network obtained using classical methods can help greatly to improve this data-based method. % Some recent works has proposed to use GP for power system operation and control problems \cite{pareek2019non,pareek2019probabilistic}.

A general GP regression, with a training data set $\mathcal{D}=\{\mathbf{x}^{i},\, \hat f (\mathbf{x}^{i})\}^N_{i=1}$ where $\hat f (\mathbf{x}^{i})=\hat y(\mathbf{x}^{i})$ is the measured function value at input $\mathbf{x}^{i} \in \mathbb{R}^n$ at the $i$-th step, is given as \cite{williams2006gaussian}:
\begin{align}\label{eq:reg}
        \hat y(\mathbf{x}^{i}) =y(\mathbf{x}^{i})+\varepsilon^{i}, \quad i=1 \dots N.
        \vspace{-1em}
\end{align}
 
In the POPF problem, vector $\mathbf{x}$ contains uncertain power injections, load demands, and renewable generations, while $\hat y$ refers to the optimal ACOPF output corresponding to the input point $\mathbf{x}$. The optimal output $\hat y$ can be optimal cost, generator dispatch, and node voltage values obtained with ACOPF solution. In \eqref{eq:reg}, $\varepsilon^{i}$ are independent and identically distributed noise variable with zero-mean, $\sigma_n$ standard deviation normal distribution. Interested reader can look into \cite{williams2006gaussian} for details of GP fundamentals and compact expressions of mean and variance values of the posterior distribution. 
The GPR establishes the relationship between input and output variables via the covariance function $k(\mathbf{x}^i,\mathbf{x}^j)$. The kernel function is selected based upon the prior understanding of the problem at hand. We use squared exponential (SE) covariance to model the POPF under the load and renewable generation uncertainties due to the smoothness of kernel. The SE covariance is given as:  
 \begin{align}\label{eq:kernel}
     k(\mathbf{x}^i,\mathbf{x}^j) = \sigma_f^2 \exp\big \{-0.5(\mathbf{x}^i-\mathbf{x}^j)^T M (\mathbf{x}^i-\mathbf{x}^j) \big \}.
         \vspace{-0.5em}
 \end{align}

 Here, $M = l^{-2}I$ is a matrix having \textit{characteristic length} ($l$) while $\sigma_f^2$ is the normalised \textit{scaling factor}. Both, $l$ and $\sigma_f$ are collectively called \textit{hyperparameters}. These \textit{hyperparameters} contain the subtle information about the function, providing interpretations on behaviour. 
 
\vspace{-0.5em}
 \section{Proposed GP-POPF}

The basic deterministic OPF problem can be expressed as: 
\vspace{-0.5em}
\begin{equation}
\begin{aligned} \label{eq:DOPF}
    \min ~ & c(\mathbf{y}) \\
     \text{s.t.} ~ & g(\mathbf{x},\mathbf{y})=0 ~; \, \, h(\mathbf{x},\mathbf{y}) \leq 0.
\end{aligned}
\vspace{-0.4em}
\end{equation}
Here, $\mathbf{y}$ is the output set including generation and voltage setpoints while $c(\cdot)$ is the associated cost function. $\mathbf{x}$ represents the input vector combining the load and renewable generations. $g(\mathbf{x},\mathbf{y})$ refers to the non-convex power balance constraint while $h(\mathbf{x},\mathbf{y})$ represents other operational inequality constraints like the upper and lower limits of the variables. With random $\mathbf{x}$, the deterministic OPF in \eqref{eq:DOPF} can be cast as Probabilistic OPF.

\color{black}
 In the proposed GP-POPF, GPR is used as an uncertainty propagation method, i.e., obtaining the output variable distribution of OPF for a given input uncertainty. For this, the first stage of the proposed GP-POPF is to construct a learning data set $\mathcal{D}_l=\{\mathbf{X},\, \mathbf{\hat Y} \}$ by solving \eqref{eq:DOPF} for $N$ different input vectors respectively. Here, $\mathbf{X} \in \mathbf{R}^{N \times n}$ with each row being one $n-dimensional$ uncertain input vector, $\mathbf{x} = [\mathbf{P}^T_r \, \mathbf{P}^T_d \, \mathbf{Q}^T_d]$ with $\mathbf{P}_r\, ,  \mathbf{P}_d \,$ and $\mathbf{Q}_d$ are uncertain renewable generation, real and reactive demand vectors. The output matrix $\mathbf{\hat Y} \in \mathbf{R}^{N \times m}$ contains columns of $m$ different output variables as optimal cost $c(\mathbf{P}^o_g)$, optimal generator set points ($\mathbf{P}^o_g \,, \mathbf{Q}^o_g$) and node voltages $\mathbf{V}^o$, corresponding to each uncertain input vector in $\mathbf{X}$. The GPR is used to learn all these output variables independently and in parallel for a better computational performance. For the construction of uncertain input set, we use uniform distribution with a box type uncertainty description as $\mathbf{x} \in [\mathbf{x}^-, \mathbf{x}^+]$. By maximizing the Log marginal likelihood \cite{williams2006gaussian}, we obtain the mapping $f_j: \mathbb{R}^n \mapsto \mathbb{R}$ for each individual output variable as $  y_j=f_j(\mathbf{x})$ for $j=1,\dots m$.
 
\begin{table}[b]
  \centering
  \vspace{-1.8em}
  \caption{Error in optimal cost parameters with $ \pm 10 \%$ load uncertainty}
  \vspace{-0.8em}
    \begin{tabular}{c|cccccc}
    System &  Penetration  &  \% Error in $\mu(cost)$ & \% Error in $\sigma(cost)$ \\
    \hline
    14-Bus & 10.42\%   & 0.000066 & 0.001700 \\
    14-Bus & 17.37\%   & 0.000124 & 0.003000 \\
    30-Bus & 8.82\%   & 0.000085 & 0.000108 \\
    30-Bus & 24.70\%  & 0.000004 & 0.001400 \\
    \hline
    \end{tabular}%
  \label{tab:cost}%
%   \vspace{-1.9em}
\end{table}% 
 
One can learn the voltage magnitude relationship, for example, at $j$-th bus, with random renewable generation vector $\mathbf{P}_r$. Then based on the function-space view of GPR (Chapter 2 \cite{williams2006gaussian}) the mean prediction of voltage magnitude output $V_j$ for an arbitrary, sample input vector $\mathbf{P}_r^s\in [\mathbf{P}_r^-, \mathbf{P}_r^+] $ is given as:
\vspace{-0.5 em}
\begin{align} \label{eq:form}
    % \bar y_{j}(\mathbf{x}_\star) = \sum^N_{i=1}\alpha^i_j k_j(\mathbf{x}_{i.},\mathbf{x}_\star).
        V_{j}(\mathbf{P}_r^s) =  k_j(\mathbf{X},\mathbf{P}_r^s)^T\boldsymbol{\alpha}_j.
    \vspace{-1.5em}
\end{align}

Here, $\boldsymbol{\alpha}_j=(K_j+\sigma_n I)^{-1} \mathbf{\hat y}_j$ with $K \in \mathbb{R}^{N \times N}$ being the covariance matrix obtained on input variables from $\mathcal{D}_l$, and $k_j(\mathbf{X},\mathbf{P}_r^s) \in \mathbb{R}^{N \times 1}$ is kernel vector. Similarly, the variance at test input vector $\mathbf{P}_r^s$ can be obtained. Here, it is important to highlight that once training is done, $\boldsymbol{\alpha}_j$ remains constant. Thus, repeated matrix inversion is not required. The form of function $ V_j(\mathbf{P}_r^s)$ in \eqref{eq:form} can be understood as an optimal linear combination approximation of ``basic kernel'' $k_j(\mathbf{X},\mathbf{P}_r^s)$. The set of $\boldsymbol{\alpha}_j$'s represent the optimal weights or coefficients that best fits the concerned function $y_j=f_j(\mathbf{x})$. For a new input point $\mathbf{P}_r^s$, one just needs to plug in such new value of $\mathbf{P}_r^s$ into \eqref{eq:form} to calculate the optimal output solution $ V_{j}(\mathbf{P}_r^s)$. Similarly, voltage-reactive power and other relationships can be learnt. 

The main advantages of the proposed GP-POPF method over existing techniques can be summarized as follows: 

\begin{itemize}
    \item At the learning stage, no assumption is made on the type of uncertainty distribution for the input vector $\mathbf{x}$.
    \item During prediction, the GP-POPF method provides output distributions for any input uncertainty distribution.
    \item Unlike other uncertainty propagation methods, GP provides variance information directly, which can be used to obtain probabilistic limits on output variations.
    \item GP is an Interepretable Machine Learning (IML) tool. Thus, \textit{hyperparameters} can be used to obtain more insights into GP-POPF such as relative output variations.
    \item Proposed GP-POPF makes no assumption in power flow equations for uncertainty propagation. Full ACOPF models, with different objectives and constraints, can be utilized to understand system behavior. % Further, different OPF models can also be learnt using proposed method. 
\end{itemize}

\vspace{-0.5em}
\section{Results and Discussion}

In this section, we present the results obtained on the IEEE 14-bus and 30-bus system \cite{babaeinejadsarookolaee2019power}. The 14-bus system has three renewable generators connected at bus 7, 9, and 14 with total of 11 uncertain load nodes. The 30-bus system has five renewable generators at 6, 9, 22, 25, and 28-th bus. All 21 load buses uncertain apparent power demand. All the renewable generators have an uncertainty of $100\%$, thus covering entire uncertain space and does not require any specification of the type of generator and power distribution. The MATPOWER ACOPF solutions (from \textit{runopf}) \cite{zimmerman2010matpower} are used to obtain OPF solutions using $\mathbf{x} = [\mathbf{P}^T_r \, \mathbf{P}^T_d \, \mathbf{Q}^T_d]$. The testing is performed with MCS on $10^4$ random input samples. As the proposed Bayesian inference-based GPR is a non-parametric method, the proposed POPF method can be used to get inference for any distribution of uncertainty once the model is trained. 

\begin{figure}[b]
    \centering
    \vspace{-2.1em}
    \includegraphics[width=\columnwidth]{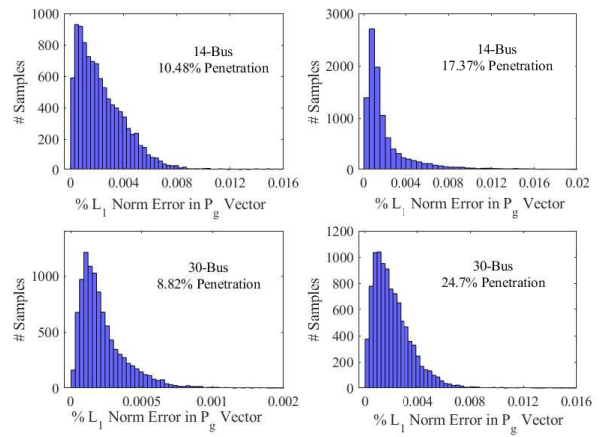}
    \vspace{-2.0em}
	\caption{$\%L_1$ error in $\mathbf{P}_g$, with $\pm 10 \%$ load uncertainty, for $10^4$ sample MCS}
    \label{fig:pgerror}
        %   \vspace{-1.5em}
\end{figure}

\begin{figure}[t]
    \centering
    \vspace{-0.5em}
    \includegraphics[width=\columnwidth]{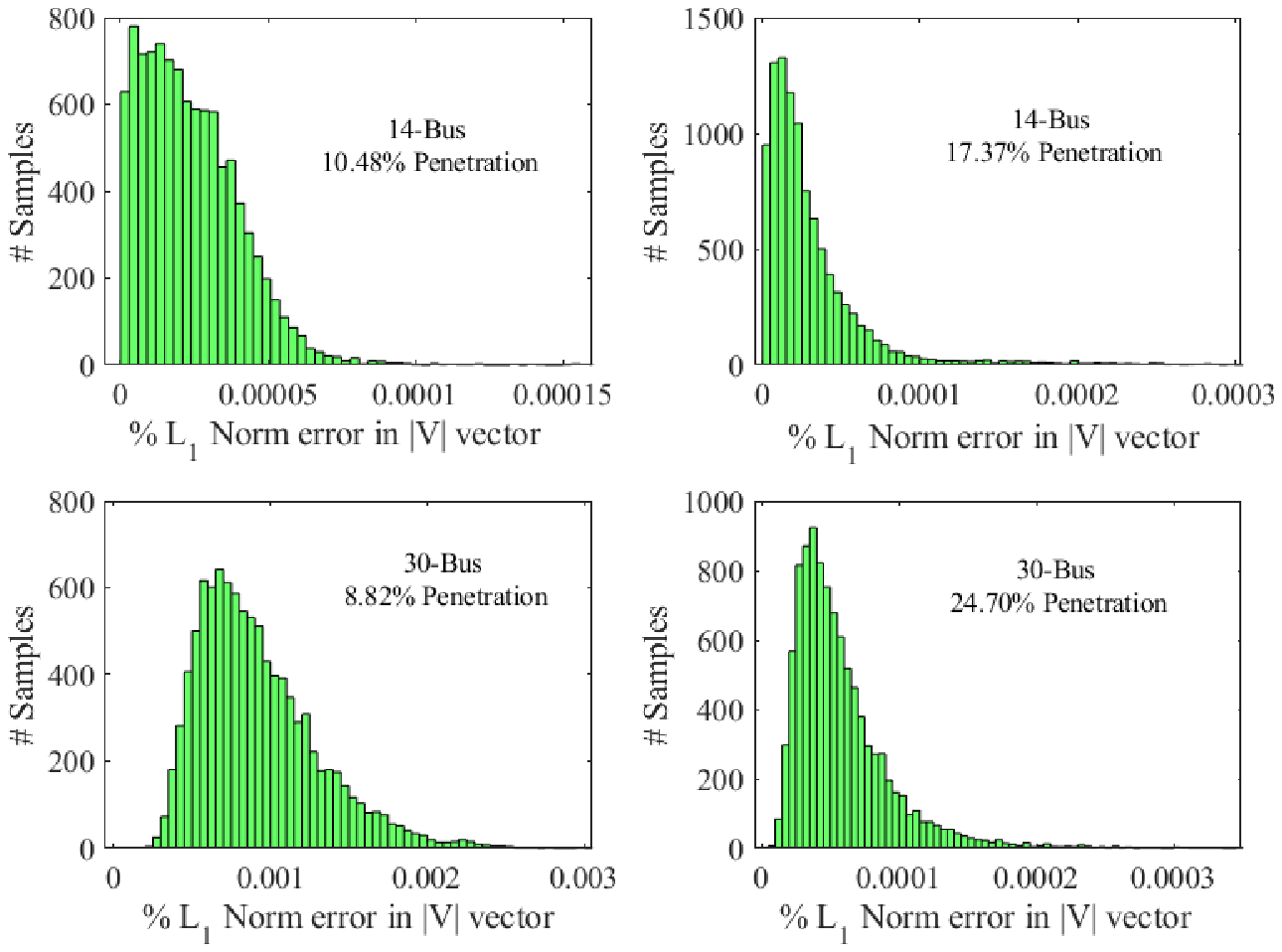}
    \vspace{-2em}
	\caption{$\%L_1$ error in $\mathbf{|V|}$, with $\pm 10 \%$ load uncertainty, for $10^4$ sample MCS}
    \label{fig:Verror}
          \vspace{-1.7em}
\end{figure}

Table \ref{tab:cost} contains the error in mean and standard deviation of optimal cost $c(\mathbf{P}^o_g)$ distribution considering different levels of renewable penetration with $\pm 10\%$ load uncertainty. The error values show that the proposed GP-POPF has been able to achieve very high accuracy in cost estimation. Further, we define $\% L_1$ error as the $L_1$ norm distance between MCS solution $\mathbf{\hat y}$ and GP-POPF solution $\mathbf{y}$ as ${\norm{\mathbf{\hat y}-\mathbf{y}}_1}/{\norm{\mathbf{\hat y}}_1}\times 100$. This error indicates the mismatch between true and estimated solution. The Fig. \ref{fig:pgerror} presents $\% L_1$ error distribution for vector $\mathbf{P}_g$ while Fig. \ref{fig:Verror} shows the error distribution corresponding to $\mathbf{V}$ for different system cases. Clearly, the mean of error distribution is close to zero, while the number of samples decreases very fast as we move to higher error bins. Further, the low values of  $\% L_1$ error establish the applicability of the proposed GP-POPF as the predicted output decision, and state variable will require very fewer adjustments in case they do not satisfy physical and operational constraints. The detailed adjustment method development is not in the scope of this letter and will be developed subsequently.

In the following, we report the computation time with unoptimized codes.
% The traditional GP may $\mathcal{O}(N^3)$ complexity in time, which $N$ being the number of training samples. However, 
As training can be done offline and in parallel, GP will not become a direct bottleneck in the proposed GP-POPF. In the simulations presented here, we require $N \leq 300$ training samples for both the systems at different levels of renewable penetration. This leads to the computation time of $4.28s$ for the 14-bus system and $ 4.67s$ for the 30-bus system to complete training and prediction over $10^4$ testing samples using parallel execution by unoptimized codes. On the contrary, the MCS takes $245.82\,s$ for 14-bus and $ 357.08\,s$ for 30-bus system for $10^4$ sample results. Further, when the proposed GP-POPF is once trained, it can be used for any number of predictions using \eqref{eq:form}. For elevating the time complexity issue, methods reviewed in \cite{liu2020gaussian} can be considered. For simulations, we have used the GPML toolbox with MATLAB 2018b on PC having Intel Xeon E5-1630v4@3.70 GHz, 16 GB RAM.

\vspace{-1.0 em}

\subsection{Observations on Interpretability: Subspace-wise Sensitivity}

\begin{figure}[t]
    \centering
    \includegraphics[width=0.85\columnwidth]{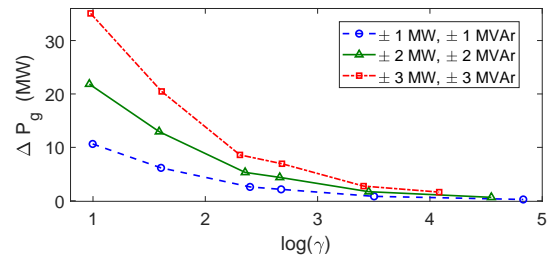}
    \vspace{-1em}
	\caption{Inverse relationship of $\Delta P_g$ with \textit{Subspace Sensitivity} $\gamma=l/\sigma_f$ for all the generators in 30-bus system at different load uncertainties.}
    \label{fig:inter}
          \vspace{-1.5em}
\end{figure}

As GP is an Interpretable Machine Learning tool, the proposed GP-POPF can be used to understand the steady-state behavior of the power system under uncertainty. One such observation is on the range of variations in generator setpoints with uncertain $\mathbf{x}$. The interpretation is targeted to check if there exists any relationship between kernel \textit{hyperparameters}, $l$ and $\sigma_f$ in \eqref{eq:kernel}, and $P_g$ variation, $\Delta P_g = \max \{P_g\} -\min \{P_g\}$, over input uncertain subspace where the $\Delta P_g$ gets affected simultaneously by the cost function, non-linearity of power balance and various inequalities in POPF. The two different \textit{hyperparameters} are indicative of different features of the function $f_j(\mathbf{x})$. The \textit{characteristic length } $l$ indicate "wiggle" lengths and higher value of $l$ means the function various slower (or less) with respect to variations in input. The $\sigma_f$ indicates the average distance of function from its mean. As both \textit{hyperparameters} are obtained together using optimization, it is important to interpret them with each other. Therefore, we define ratio of these hyperparameters as $\gamma=l/\sigma_f$ which indicates the normalised variations in $f_j(\mathbf{x})$ within the input subspace. The lower value of $\gamma$ means that the function is highly non-linear and has large variations from its mean. Figure \ref{fig:inter} depicts observation establishing this interpretation which shows an inverse relationship between $\gamma$ and $\Delta P_g $, for all generators. This means that a generator with less non-linearity i.e., higher $\gamma$, will have fewer variations in $P_g$. % Thus, $\gamma$ depicts the normalised variation in $y_j$ with respect to $\mathbf{x} \in [\mathbf{x}^-, \mathbf{x}^+]$. The proposed GP-POPF provides this information without any extra calculation. 
$\gamma$ can be considered as the output sensitivity over an input subspace as \textit{hyperparameters} are obtained by optimization over the entire input subspace $\mathbf{x} \in [\mathbf{x}^-, \mathbf{x}^+]$. This is much different than the traditional derivative-based sensitivity calculations which are valid only at that operating point. We call it \textit{Subspace-wise Sensitivity} $\gamma$ and will be explored in the future.

\color{black}
\vspace{-0.5em}

\section{Conclusion}
Different from existing analytical methods, the proposed GP-POPF method does not rely on uncertainty information and linearization assumptions on the power flow. Compared to data-based methods, the proposed method does not require extensive training samples of POPF solutions, thus reducing computation time. The Bayesian inference-based GP is used for uncertainty propagation and allows one to explore a new path for interpretability for the first time through the concept of \textit{Subspace-wise Sensitivity}. The letter opens up the possibility of exploring interpretable models for uncertainty handling in power system operation and control applications. 

\vspace{-0.5em}
\bibliographystyle{IEEEtran}
\bibliography{main}
\end{document}